\documentclass[aps,prb,twocolumn,superscriptaddress]{revtex4-1}
\usepackage{graphicx}
\usepackage{picture}
\usepackage{xcolor}
\usepackage{color}
\usepackage{tikz}

\begin{document}

\title{Bridging the Gap Between the Transient and the Steady State of a Nonequilibrium Quantum System}

\author{Herbert F. Fotso}
\author{Eric Dohner}
\affiliation{Department of Physics, University at Albany (SUNY), Albany, New York 12222, USA}
\author{Alexander Kemper}
\affiliation{Department of Physics, North Carolina State University, Raleigh, NC, 27695, USA}
\author{James K. Freericks}
\affiliation{Department of Physics, Georgetown University, 37th and O Sts. NW, Washington, DC 20057, USA}

\begin{abstract}

Many-body quantum systems in nonequilibrium remain one of the frontiers of many-body physics. While there has been significant advances in describing the short-time evolution of these systems using a variety of different numerical algorithms, it has been quite difficult to evolve a system from an equilibrium state prior to the application of a driving field, to the long-time steady (or periodically oscillating) state. These dynamics are complex: the retarded quantities tend to approach their long-time limit much faster than the lesser (or greater) quantities. Recent work on strongly correlated electrons in DC electric fields illustrated that the system 
may evolve through successive quasi-thermal states obeying an effective fluctuation-dissipation theorem in time. We demonstrate an extrapolation scheme that uses the short-time transient calculation to obtain the retarded quantities and to extract how the lesser/greater quantities vary with time and then extend the numerical solutions all the way to the steady state, with minimal additional computational cost. Our approach focuses on extrapolating the electronic self-energy and then employing that to determine the Green's function and various experimentally relevant expectation values.

\end{abstract}

\maketitle


\section{Introduction}
\label{sec: Introduction}

When a quantum system is driven away from equilibrium, its subsequent relaxation can unfold in different stages, starting with an early transient and ending with a steady state that may or may not be thermal. The dynamics of the nonequilibrium system in these different stages has been the subject of intense activity in many subfields of physics including strongly correlated quantum systems.

Dynamical mean-field theory (DMFT)\cite{DMFT, DMFT_2, DMFT_3, DMFT_FK} and its cluster extensions\cite{DCA_review, DCA_1, DCA_2, DCA_3, CDMFT}, have been fairly successful methods within the toolkit to study strongly correlated electronic systems in equilibrium. They have been extended to the nonequilibrium situation and used with some success\cite{DMFT_noneq, FK_NonEq_DMFT08, DMFT_noneq_Aoki, NonEq_DCA}. These methods map the lattice problem onto an impurity (in the case of DMFT), or a cluster (for cluster methods) embedded into a self-consistently determined bath. The effectiveness of the formalism then relies on the ability to efficiently solve this correlated impurity/cluster problem. For the nonequilibrium DMFT solution in particular, several approaches have been employed in this respect. These include continuous time Quantum Monte Carlo (CTQMC)\cite{CTQMC_NonEq}, exact diagonalization\cite{KrylovPotthoff, Arrigoni_von_der_linden}  and various perturbative methods\cite{pertubationTheory_NonEq, pertubationTheory_NonEq_2}. These solvers are however constrained either by the sign problem or the exponential growth of the Hilbert space with increasing number of degrees of freedom that are already troublesome in equilibrium but  are exacerbated in nonequilibrium. As a result, the impurity solvers have been limited in solutions of nonequilibrium problems to relatively short time transients. This problem is particularly acute for perturbative solvers, as perturbation theory is equivalent to a Taylor series in time and hence becomes increasingly more difficult to extend to the long-time regime.

The nonequilibrium DMFT approach has been used to study the transient relaxation of isolated interacting quantum systems initially in equilibrium and suddenly placed under the influence a DC electric field \cite{DMFT_noneq, FK_NonEq_DMFT08,    thermalization}. Additionally, a nonequilibrium DMFT solution can be formulated for the system after it has undergone its transient and settled into its steady state\cite{steadyState1, steadyState_Aoki}. However, these results shed no light on the final evolution of this transient state to the steady state. An earlier analysis of the transient identified a set of scenarios that the system follows in its relaxation towards its steady state depending on its specific parameters\cite{thermalization}.

Among the identified relaxation scenarios, both the Hubbard and the Falicov-Kimball model were found to sometimes follow a monotonic thermalization process. Here, after an initial nontrivial short period in the transient, governed primarily by the relaxation of the retarded quantities, the system subsequently evolves towards a thermal infinite temperature steady state by going through a succession of quasi-thermal states; in other words, the retarded quantities remain fixed, and the populations of the electrons evolve toward their steady state through a well-characterized behavior that can be predicted for future times. In this paper, we consider this monotonic thermalization scenario for a system described in equilibrium by the Falicov-Kimball model~\cite{FalicovKimball, AtesZiegler} which can be viewed as modeling a Fermi-Fermi heavy-light mixture. It can also be obtained from the Hubbard model by fixing one of the spin species on the lattice (heavy), while allowing the other to hop (light). We demonstrate an extrapolation scheme that allows us to bridge the gap between the transient and the steady state characterization of the field-driven system.
 
Since the lattice Green's function in the DMFT formalism is defined by the local self-energy, the extrapolation scheme is based on an analysis of the self-energies. This analysis enables us to extend the self-energies beyond the initial transient calculation and thus to be able to obtain the lattice Green's function and other physical quantities well beyond the initial transient calculation with little added computational cost.

The paper is organized as follows. In section (\ref{sec:NonEqDMFT}), we present a brief overview of the nonequilibrium DMFT formalism for the transient and for the steady state. In section (\ref{sec:NonEqSelfEnergy}), we present the self-energies for the equilibrium and steady state solutions. In section (\ref{sec:timeExtension}), we discuss the transient self-energy, our extrapolation scheme to extend the self-energies beyond the transient maximum time and how it produces observables at arbitrarily longer times.

\section{Nonequilibrium Dynamical Mean Field Theory}
\label{sec:NonEqDMFT}

We consider a system, initially in equilibrium, described by the Falicov-Kimball model. Before the application of the electric field, it is described by the Hamiltonian:
\begin{eqnarray}
{\mathcal H}_{eq}  &=&  -  \frac{1}{2\sqrt{d}} \sum_{\langle ij\rangle} t^*_{ij} \left( c^{\dagger}_{i}c_{j}^{\phantom\dagger} + c^{\dagger}_{j}c_{i}^{\phantom\dagger}
 \right) \nonumber \\
 &-&  \mu \sum_{i}  c^{\dagger}_{i}c_{i}^{\phantom\dagger}  +  U\sum_{i}w_{i}c^{\dagger}_{i}c_{i}^{\phantom\dagger};
\label{eq:HamiltonianFK}
\end{eqnarray}
the asterisk indicates a renormalized hopping for the infinite-dimensional limit and is not a complex conjugate---in equilibrium, we assume the hopping is real.

In the Hamiltonian, $\langle ij \rangle$ represents the nearest-neighbor pair of sites $i$ and $j$; $J_{ij} = t_{ij}^*/2\sqrt{d}$ is the hopping integral between nearest-neighbor sites $i$ and $j$ and $t_{ij}^* = t^*$, the effective hopping integral, is used as the energy unit. $d$ is the spatial dimension of the system and we work in the limit $d \to \infty$. The operators $c^{\dagger}_{i}$ $(c_{i})$ create (destroy) an electron at site $i$. $\mu$ is the conduction electron chemical potential. $U$ is the on-site repulsion for doubly occupied sites and $w_{i}=0, 1$ is the occupation number operator of the localized fermions at site $i$. We study this model on an infinite-dimensional hypercubic lattice at half-filling, when there are as many conduction electrons as there are localized electrons and the total number of electrons is equal to the total number of  lattice sites. This system is driven out of equilibrium by applying an electric field at a time $t_0$ along the diagonal of the hypercubic lattice, $\mathbf{E}=E(1,1,1,, \dots)$, and subsequently keeping this electric field constant.

\begin{figure}[htbp]
\begin{center}
\includegraphics*[width=7.0cm, height=4.0cm]{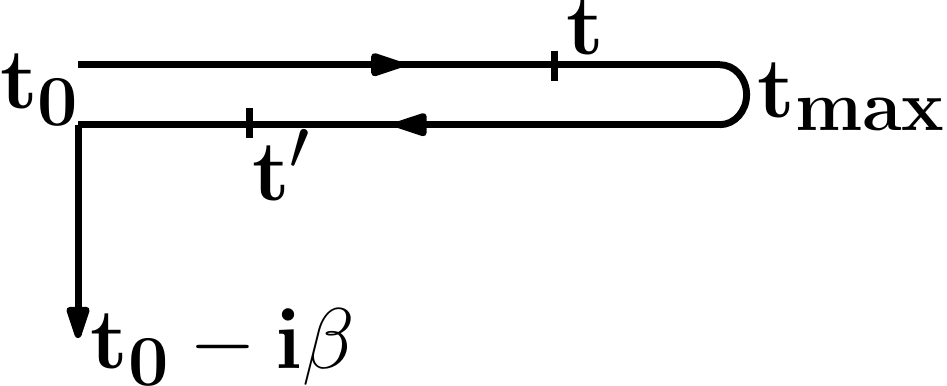}
\caption{The Kadanoff-Baym-Keldysh contour. The arrows indicate the direction of the time evolution. In this figure, $t$ occurs before $t'$ on the contour.
} 
\label{fig:KeldyshContour}
\end{center}
\end{figure}

The nonequilibrium many-body formalism is formulated on the Keldysh contour whereby the system is evolved from the distant past to the time of physical interest and then back to the distant past again\cite{Keldysh64_65}. This follows simply from the Heisenberg time evolution, which involves a similarity transformation of the operators with respect to the time-evolution operators. One operator moves forwards in time, while its Hermitian conjugate evolves backwards in time. It involves several types of two-time Green's functions among which  $G^<(t,t')$  (the lesser),  $G^>(t,t')$ (the greater), and $G^R(t,t')$ (the retarded) Green's functions.
In the context of a system initially in equilibrium at an initial temperature $T=1/\beta$, a vertical spur of imaginary times is added to the Keldysh contour resulting in the so called Kadanoff-Baym-Keldysh contour\cite{Keldysh64_65, BaymKadanoff62} illustrated in Fig.~\ref{fig:KeldyshContour}. To the previous types of Green's functions in the formalism, one should then add the Matsubara Green's function and the mixed time Green's functions, where one of the times is on either horizontal branch of real times, while the other is on the vertical branch of imaginary time.

One can immediately see that the formalism can be built around the contour-ordered Green's function (defined for each time on the contour) and all other Green's functions can be extracted from it by restricting the times to different regions on the contour.  Various identities exist between the different Green's functions and they are not all independent. All the information is contained in the retarded and the lesser Green's functions. The former defines the states in the system, while the latter characterizes how those states are occupied. The contour-ordered Green's function has time ordering performed with respect to time advance along the contour and is defined via:
\begin{equation}
    G^c_{k, \sigma}(t, t')  =  \theta_c(t, t')G^>_{k, \sigma}(t, t') + \theta_c(t', t)G^<_{k, \sigma}(t, t'). \;\;\;\;\label{eq:GContour} 
\end{equation}
Here, $\theta_c(t,t')$ is the contour-ordered Heaviside function, which orders time with respect to the contour: it is equal to $1$ if $t$ is ahead of $t'$ on the contour and is equal to $0$ if $t$ is behind $t'$ on the contour. The lesser and greater  Green's functions are defined by operator averages in the Heisenberg representation:
\begin{eqnarray}
 G^<_{k, \sigma}(t, t') & = &  i\langle c^{\dagger}_{k \sigma}(t') c_{k \sigma}^{\phantom\dagger}(t)\rangle, \label{eq:GLesser}  \\
 G^>_{k, \sigma}(t, t') & = & -i\langle c_{k \sigma}^{\phantom\dagger}(t) c^{\dagger}_{k\sigma}(t')\rangle. \label{eq:GGreater}  
\end{eqnarray}
 From these, we can construct the retarded and advanced Green's functions from
 \begin{eqnarray}
 G^R_{k, \sigma}(t,t') & = & -i\theta(t-t')\langle \{c_{k\sigma}^{\phantom\dagger}(t), c^{\dagger}_{k\sigma}(t')\}\rangle \label{eq:GRetarded}. \\
 G^A_{k, \sigma}(t, t') & = & i\theta(t'-t)\langle \{c_{k\sigma}^{\phantom\dagger}(t), c^{\dagger}_{k\sigma}(t')\}\rangle.\label{eq:GAdvanced}
\end{eqnarray}
In the above expressions, $c^{\dagger}_{k \sigma}(t)$ and $c_{k \sigma}(t)$ are, respectively, the Heisenberg representation of the creation and the destruction operators for an electron of momentum $k$ and spin $\sigma$ at time $t$; $\theta$ is now the ordinary Heaviside function, {\it i.~e.},~$\theta(t-t') = 0$ if $t<t'$ and $\theta(t, t') = 1$ otherwise; $\{\mathcal{A}, \mathcal{B}\}$ is the anticommutator of operators $\mathcal{A}$ and $\mathcal{B}$. The symbol $\langle \mathcal{O} \rangle$ is the expectation value taken of the operator $\mathcal{O}$ with respect to the initial thermal state:
\begin{equation}
    \langle \mathcal{O} \rangle=\frac{{\rm Tr} e^{-\beta\mathcal{H}(t_{\rm min})}\mathcal{O}}{{\rm Tr} e^{-\beta\mathcal{H}(t_{\rm min})}},
\end{equation}
where $\mathcal{H}(t_{\rm min}) = {\mathcal H}_{eq}$ is the initial Hamiltonian before the field is turned on.

The nonequilibrium many-body formalism is similar to that of the equilibrium problem. For a given Hamiltonian, the contour-ordered Green's function is related to the noninteracting Green's function by the Dyson equation:
\begin{eqnarray}
G_{k\sigma} &(t,& t') = G^0_{k\sigma}(t, t') \nonumber \\
&+& \int_c d\bar{t} \;\int_c d\bar{t'} G^0_{k\sigma}(t,\bar{t}) \Sigma_{\sigma}(\bar{t} ,\bar{t'})G_{k\sigma} (\bar{t'}, t'),
\label{eq:FullG}
\end{eqnarray}
where the integrals each range over the entire Kadanoff-Baym-Keldysh contour. 
$G^0_{k\sigma}(t,t')$ is the nonequilibrium noninteracting Green's function. Before the field is turned on, this noninteracting system is described by the Hamiltonian: 
\begin{equation}
\mathcal{H}_0 = \sum_{k\sigma} ( \epsilon_k - \mu ) c_{ k\sigma }^{\dagger}c_{k \sigma }^{\phantom\dagger}.
\label{eq:HamiltonianNonInt}
\end{equation}
Here, $\epsilon_k$ is the band energy for the lattice electrons. On a hypercubic lattice in infinite dimensions, it is given by:
\begin{equation}
\epsilon_k = - \lim_{d \to \infty} \frac{t^*}{\sqrt {d}} \sum_{i=1}^{d} \mathrm{cos}( k_i a),
\end{equation}
where $a$ is the lattice spacing (and is set equal to 1). 
The electric field is introduced through a Peierls' substitution that changes the hopping amplitude to one with a time-dependent phase:
\begin{equation}
J_{ij} \to J_{ij} \mathrm{exp}\left[-i\int_{\textbf{R}_i}^{\textbf{R}_j} \textbf{A}(\textbf{r},t) \cdot d\textbf{r} \right].
\end{equation}
For our present problem of a constant electric field along the diagonal, switched on at time $t=0$ and subsequently kept constant, in a vector-potential-only gauge, $\mathbf{A}=-\mathbf{E}\theta(t)$. We have set $e, \; \hbar \; \mathrm{and} \; c$ equal to $1$.

In these conditions, the noninteracting Green's function is given by\cite{noninteracting_Turkowski}: 
\begin{eqnarray}
G^{0}_{k\sigma}(t, t') &=& -i \left[ \theta_c(t, t') - f(\epsilon_k -\mu) \right]  \mathrm{e}^{i \mu (t-t') } \nonumber \\
&\times& \mathrm{exp}\left[  -i\int_{t'}^t d\bar{t} \left\{ \left. \bigg ( \mathrm{\theta}(-\bar{t})   + \mathrm{\theta}(\bar{t}) \mathrm{cos} \left( E \bar{t} \right) \right. \bigg )\epsilon_k \right. \right. \nonumber\\
&-& \left. \left. \mathrm{\theta}(\bar{t}) \bar{\epsilon}_k\, \mathrm{sin} \left( E \bar{t} \right) \right. \bigg \} \right. \Bigg ].
\label{eq: NonIntG}
\end{eqnarray}
$\bar{\epsilon}_k=  \lim_{d \to \infty} \frac{t^*}{\sqrt {d}} \sum_{i=1}^{d} \mathrm{sin}( k_i a)$ is the band velocity.

\begin{figure}[htbp]
\begin{center}
\includegraphics*[width=8.0cm, height=8.0cm]{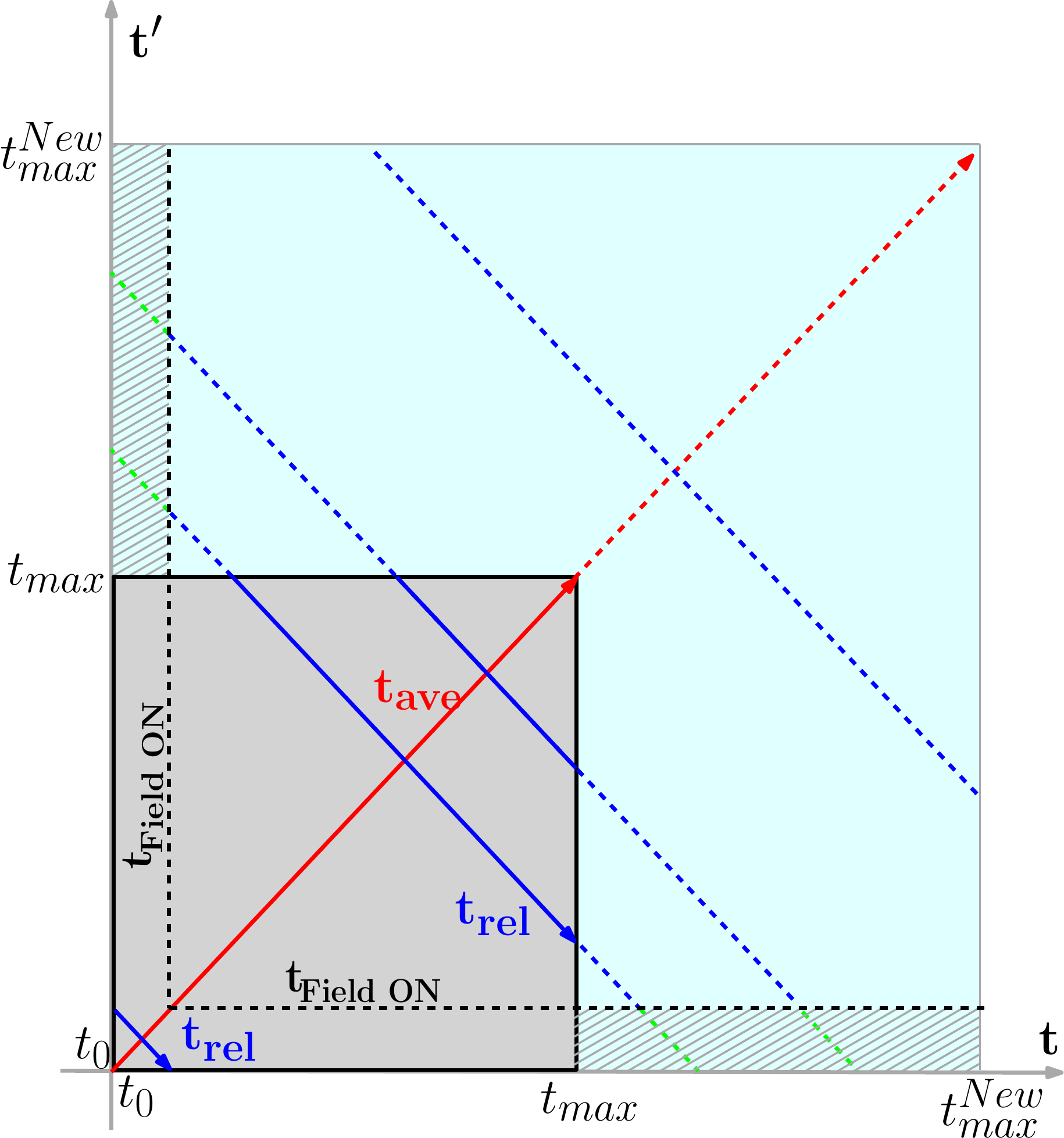}
\caption{(Color online) Schematic representation of the time coordinates on the Keldysh contour and its transformation into the Wigner coordinates. The gray area represents the short time transient calculation. The black dotted line represents the time at which the electric field is turned on. The red line represents the average time axis; To each average time corresponds a set of relative times represented by the blue lines. The graph also illustrates, in the light blue area and in the form of the dashed red and blue lines, the extension of the data beyond the initial transient calculation.  The shaded area with the dashed green lines represents time extension for data where either $t$ or $t'$ has the field off. 
} 
\label{fig:Keldysh2Wigner}
\end{center}
\end{figure}

In the DMFT formalism, the lattice problem is mapped onto an impurity problem and the self-energy is local in momentum space:
\begin{equation}
 \Sigma_{k\sigma}(t,t') = \Sigma_{\sigma}(t,t').
\end{equation}
The computational complexity of the formalism then depends on the complexity of the solution to this impurity problem. In general, no viable exact methods exist, except for the case of the Falicov-Kimball model, which we examine here.

\subsection{Transient}

As mentioned above, different methods have been implemented for the nonequilibrium problem with varied levels of success and solving this impurity problem is typically the bottleneck of nonequilibrium DMFT solutions. For the model at hand however, one can write an exact expression for the impurity Green's function:
\begin{eqnarray}
 &~&G_{imp}(t,t') = (1-\langle w \rangle) \left[ \left( i\partial_t + \mu \right) \delta_c(t,t')  - \Lambda(t,t') \right]^{-1} \nonumber \\
              &~&~~~~~~~+ \langle w \rangle \left[ \left( i\partial_t + \mu -U \right) \delta_c(t,t')  - \Lambda(t,t') \right]^{-1},
\label{eq:impurityG}               
\end{eqnarray}
where $\langle w \rangle = \sum_i \langle w_i \rangle /N$ is the initial (equilibrium) filling of the localized fermions (equal to $0.5$ at half-filling), $\Lambda(t,t')$  is the hybridization of the impurity to the dynamical mean field and $\delta_c(t,t')$ is a generalization  of the Dirac delta function onto the contour.

This nonequilibrium problem involves continuous matrix operators defined on the contour. To solve it, we must discretize the problem on a finite-size contour extending from some initial time $t_{min}$ to a final time $t_{max}$ with a finite $\Delta t$ time step. The problem is solved for three different values of $\Delta t$ and the solution is quadratically extrapolated to the $\Delta t\to 0$ limit\cite{FK_NonEq_DMFT08}. To analyze the behavior of the system, it is useful to switch representations from the $(t, t')$ time coordinates to the average and relative Wigner time coordinates $(t_{ave}, t_{rel})$, where $t_{ave} = (t+t')/2$ is the average time, while $t_{rel} = t- t'$ is the relative time, as illustrated in Fig.(\ref{fig:Keldysh2Wigner}). One may then view $t_{ave}$ as an effective time for observing the dynamics of the system; while Fourier transforming with respect to $t_{rel}$ produces frequency-dependent quantities. This approach is adopted for the solutions discussed below. Despite the fact that we can write an exact expression for the Green's function, the solutions are constrained by computational requirements both in computer time and memory\cite{FK_NonEq_DMFT08}. Thus, while the solution enables a characterization of the early time dynamics\cite{thermalization, NoneqFDT_Frontiers}, it usually cannot reach the steady state.

\subsection{Steady State}

One can alternatively formulate the nonequilibrium DMFT formalism at $t_{ave} \to \infty$, when all transient response is gone and the system is in a steady state. In this context, a formalism inspired by the Floquet method has been implemented for the steady state of our field-driven system\cite{Floquet_Shirley, Floquet_Sambe, steadyState1, steadyState2, steadyState_Aoki}. Along with a formulation of the Dyson equation in frequency space, an expression of the frequency-dependent steady-state noninteracting Green's function for a given value of the electric field is obtained. To complete the formalism, for the system of interest, which is described in equilibrium by the Hamiltonian of Eq.(\ref{eq:HamiltonianFK}), the dressed Green's function is related to the noninteracting Green's function via
\begin{equation}
G(\omega) = (1-w_1)G^0(\omega) + \frac{w_1}{G^{0^{-1}}(\omega) - U}.
\end{equation}
Here, $G^0$ is the frequency space nonequilibrium noninteracting Green's function at late average times ($t_{ave} \to \infty$) in a field. The frequency is taken in an interval determined by the Bloch frequency and a matrix structure in the Dyson equation couples each frequency to all frequencies shifted by integer multiples of the Bloch frequency.  This nonequilibrium steady state DMFT formalism allows the calculation of the density of states and related quantities once the system has undergone the transient relaxation.

\begin{figure}[htbp]
\begin{center}
\includegraphics*[width=8.50cm, height=8.50cm]{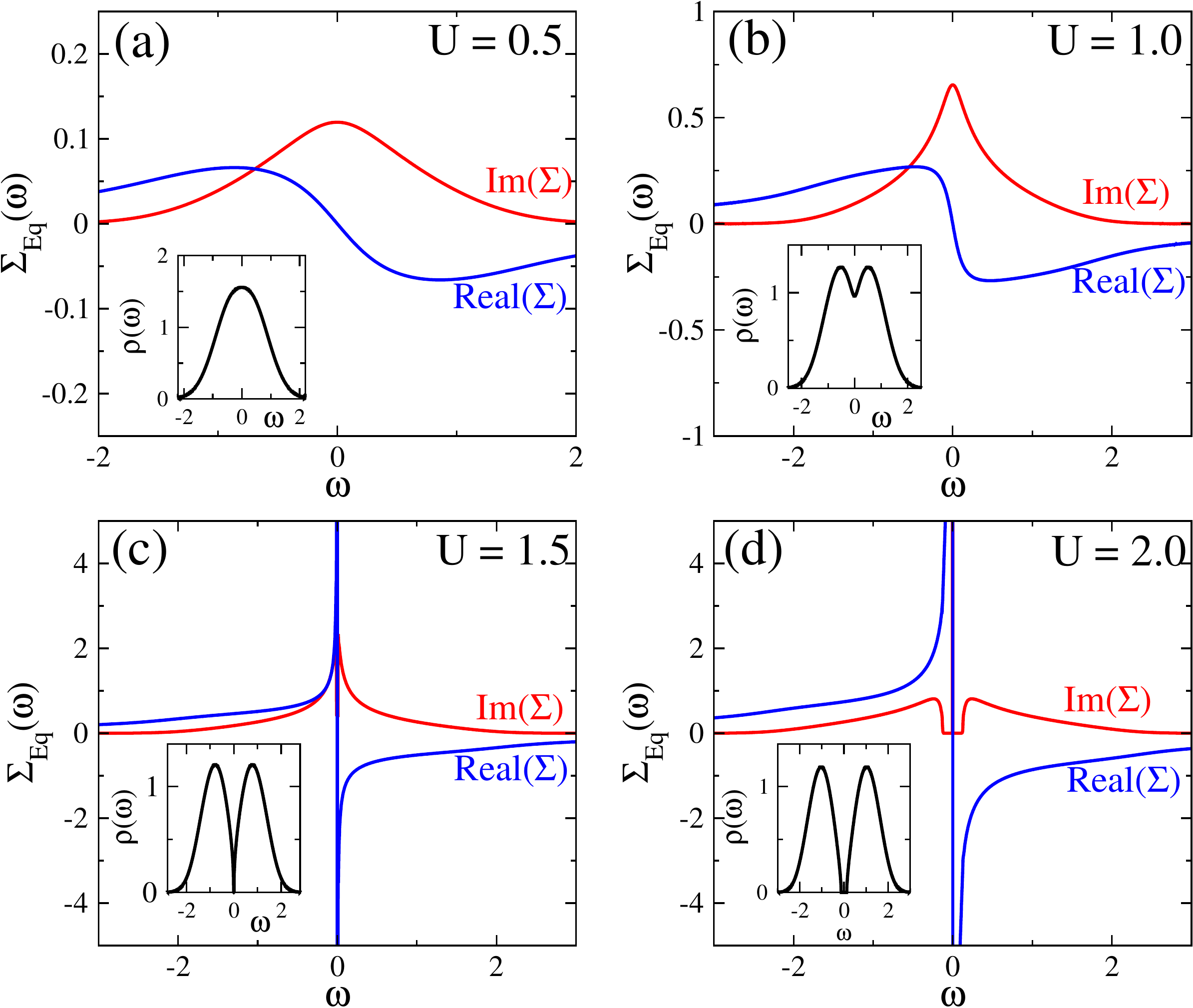}
\caption{(Color online) Real(blue) and imaginary(red) parts of the equilibrium retarded self-energy as a function of frequency for $U=0.5$ (a), $U=1.0$ (b), $U=1.5$ (c),  $U=2.0$ (d). The inset shows the corresponding density of states.
Note that the Hartree term has been subtracted out of the real part of the self-energy.
} 
\label{fig:SigmaRetardedEq_W}
\end{center}
\end{figure}

\section{Nonequilibrium Steady State and Equilibrium DMFT Self-Energy}
\label{sec:NonEqSelfEnergy}

\begin{figure}[htbp]
\begin{center}
 \includegraphics*[width=8.50cm, height=8.0cm]{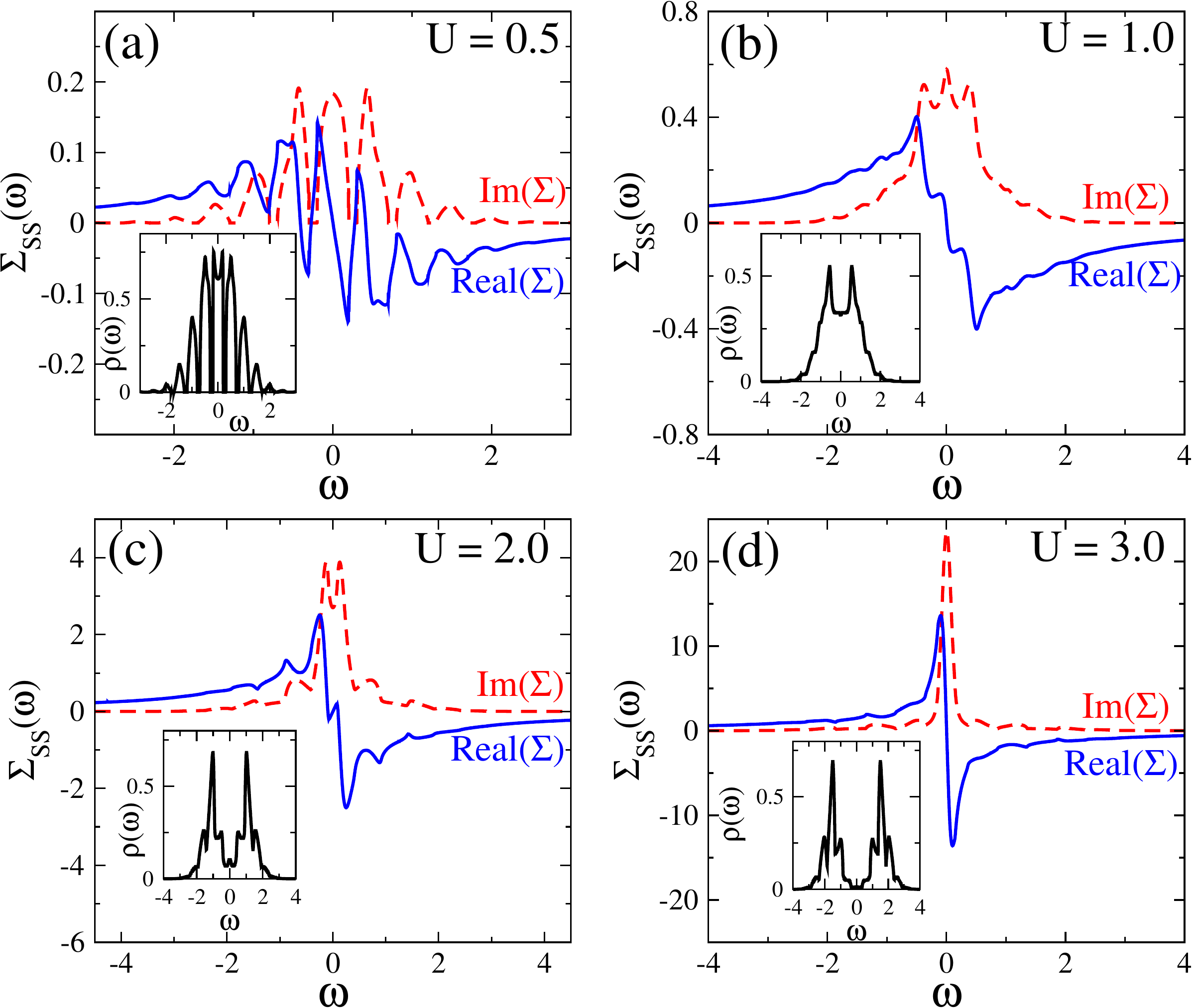}
\caption{(Color online) Real~(blue full line) and imaginary~(red dashed line) parts of the steady state retarded self-energy as a function of frequency for $E=0.5$ and $U=0.5$(a), $U=1.0$(b), $U=2.0$(c), and $U=3.0$(d). The inset of each graph shows the corresponding density of states.  Note that the Hartree term has been subtracted out of the real part of the self-energy.
} 
\label{fig:SigmaRetardedSS_W_1}
\end{center}
\end{figure}

\begin{figure}[htbp]
\begin{center}
 \includegraphics*[width=8.50cm, height=8.0cm]{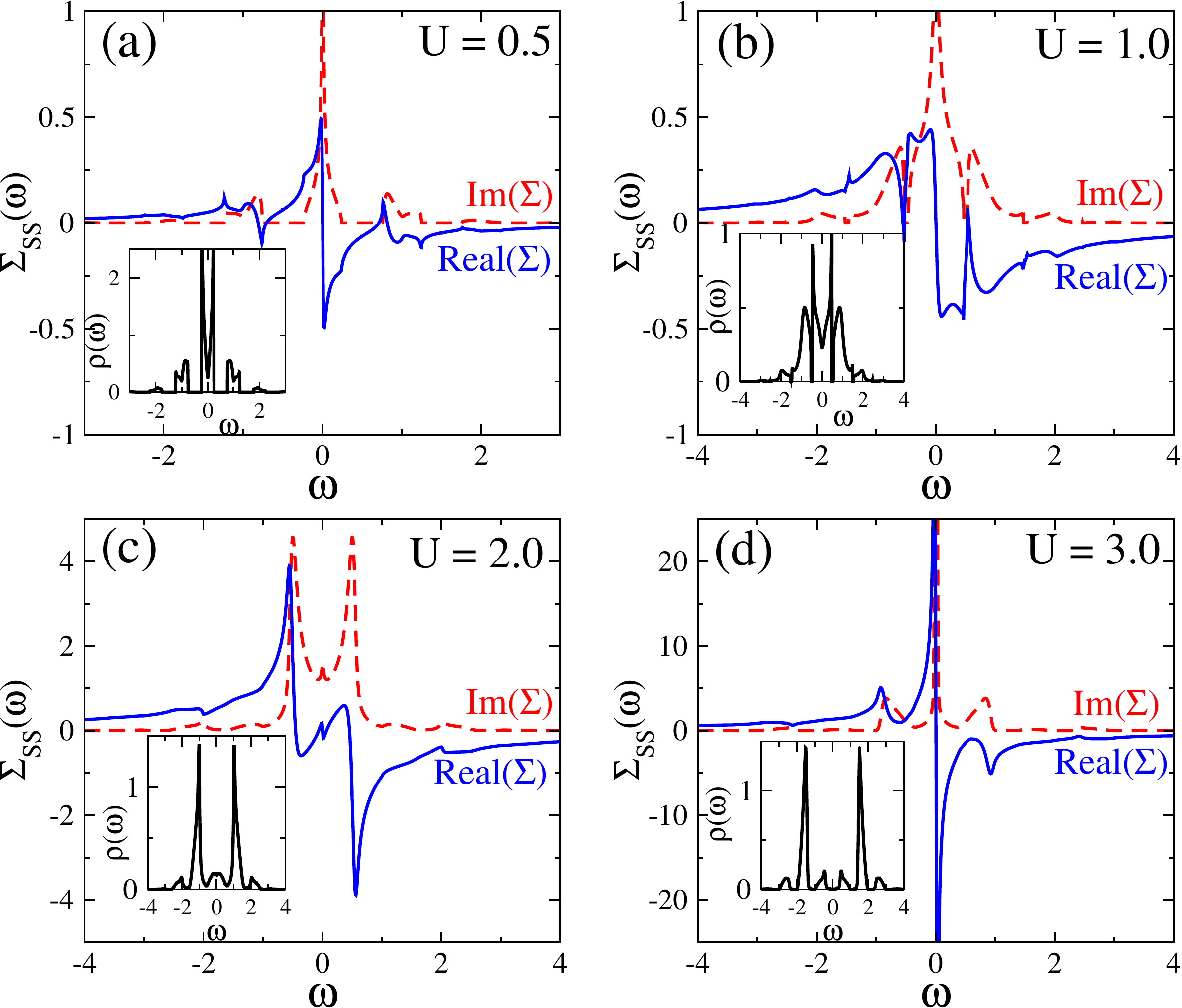}
\caption{(Color online) Real~(blue full line) and imaginary~(red dashed line) parts of the steady state retarded self-energy as a function of frequency for $E=1.0$ and $U=0.5$(a), $U=1.0$(b), $U=2.0$(c), and $U=3.0$(d). The inset of each graph shows the corresponding density of states.  Note that the Hartree term has been subtracted out of the real part of the self-energy. 
} 
\label{fig:SigmaRetardedSS_W_2}
\end{center}
\end{figure}

Within the DMFT construction, the local self-energy is necessary for the calculation of lattice Green's function and other related quantities. Thus, for our purpose of extending the characterization of the system beyond the transient calculation based on its relaxation scenario, we will focus on an analysis of the self-energies.

It is useful to start by revisiting the equilibrium self-energy results pictured in Fig.(\ref{fig:SigmaRetardedEq_W}). This figure shows the real (full blue line) and imaginary (dashed red line) parts of the equilibrium retarded self-energy as a function of frequency for $U=0.5$ (a), $U=1.0$ (b), $U=1.5$ (c),  and $U=2.0$ (d). The inset shows the corresponding density of states.  Since the noninteracting density of states for the noninteracting system on an infinite-dimensional hypercubic lattice is a Gaussian that has infinite bandwidth, the formation of a gap in the interacting system density of states corresponds to the development of a pole in the self-energy\cite{selfEnergyPoles_PRL2004, DMFT_FK}. This gap increases as the interaction strength is further increased.

Next, we present the steady state self-energies for the field driven system in Fig.(\ref{fig:SigmaRetardedSS_W_1}). The full blue line shows the real part, while the dashed red line shows the imaginary part of the retarded self-energy for $E=0.5$ and $U=0.5$(a), $U=1.0$(b), $U=2.0$(c), and $U=3.0$(c). The inset of each graph shows the corresponding density of states. The picture in this case is clearly much more complex than for the equilibrium system. The self-energy shows signatures of the presence of Wannier-Stark ladders in the density of states for the field-driven system in the case of weak interactions\cite{Wannier}. It is accordingly modified when the interaction is gradually increased. First the width of the peaks increases with $U$ until they eventually merge before ultimately giving rise to a modified spectrum that clearly features the development of in-gap states\cite{poles}. Fig.~(\ref{fig:SigmaRetardedSS_W_2}) presents the self-energies for the same Coulomb interactions and for $E=1$.

\section{From Transient to Steady State}
\label{sec:timeExtension}

The transient nonequilibrium DMFT calculation allows the characterization of the system immediately after the field is turned on. We consider here that it is initially in equilibrium at temperature $ T = 0.1$, from the earliest simulation time $t_{min}= 0$ until time $t_0 =5$ when the electric field is turned on. The system is then tracked through its dynamics until the latest simulation time $t_{max} = 40$. Here, we revisit the case of the monotonic thermalization where the system, after an early non-trivial evolution, proceeds towards an infinite temperature thermal state through a succession of quasi-thermal states. We will particularly consider the monotonic thermalization scenario for the system with electric field $E=0.5$, and interaction strength $U=1.5$~\cite{thermalization}. Details in the calculated nonequilibrium dynamics will provide an avenue for extending the characterization of the system beyond its initial transient calculation.

In a previous study~\cite{thermalization}, it was shown that the retarded Green's function adopts its steady-state value immediately after the electric field is switched on and is only constrained by causality. In addition, in the particle-hole symmetric case of interest in this work, one can show that the local retarded and lesser Green's functions are related by:
\begin{equation}
    G^R(t,t') = \theta(t-t')\left[-G^<(t,t')^* + G^<(t,t')\right]
    \label{eq:retardedLesserParticleHole}
\end{equation}
which means that the imaginary part of the lesser Green's function as a function of time is given by half of that of the retarded Green's function. Furthermore, the real part of the lesser Green's function vanishes in the infinite temperature steady state. 

\begin{figure}[htbp]
\begin{center}
  \includegraphics*[width=8.50cm,height=7.20cm]{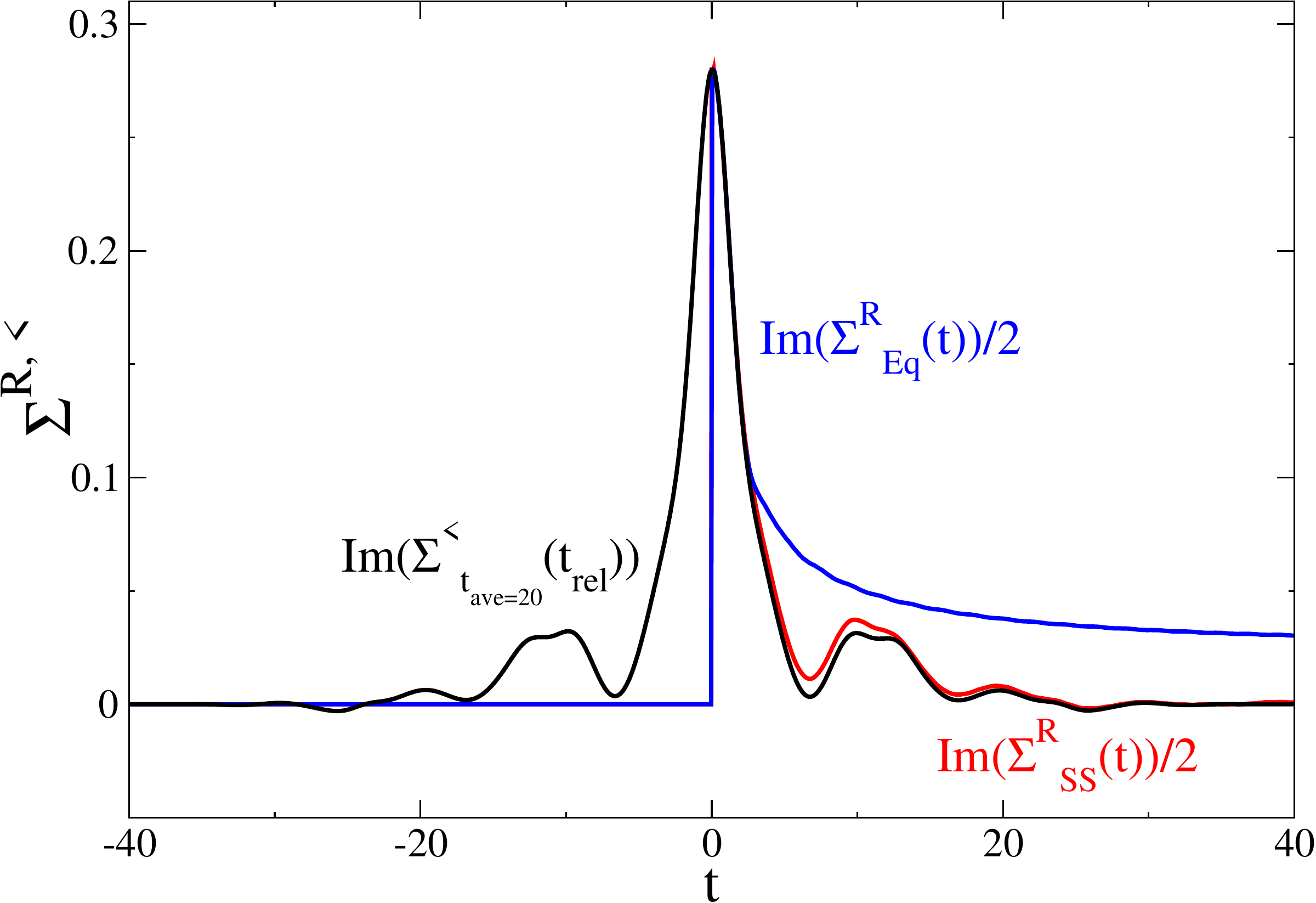}
\caption{(Color online) Imaginary part of the transient lesser self-energy at $t_{ave}=20$ and half of the imaginary part of the steady-state retarded self-energy for $E=0.5$ and $U=1.5$. The transient adopts its steady state value immediately after the field is turned on and is only constrained by causality. The imaginary part of the lesser self-energy (black) is half of that of the retarded self-energy (red) which is well matched by the steady state value despite small numerical discrepancies. The imaginary part of the equilibrium retarded self-energy as a function of time for $U=1.5$ is also shown (blue).
} 
\label{fig:SigmaRetardedSS_Eq_t}
\end{center}
\end{figure}

\begin{figure}[htbp]
\begin{center}
\includegraphics*[width=8.50cm, height=7.50cm]{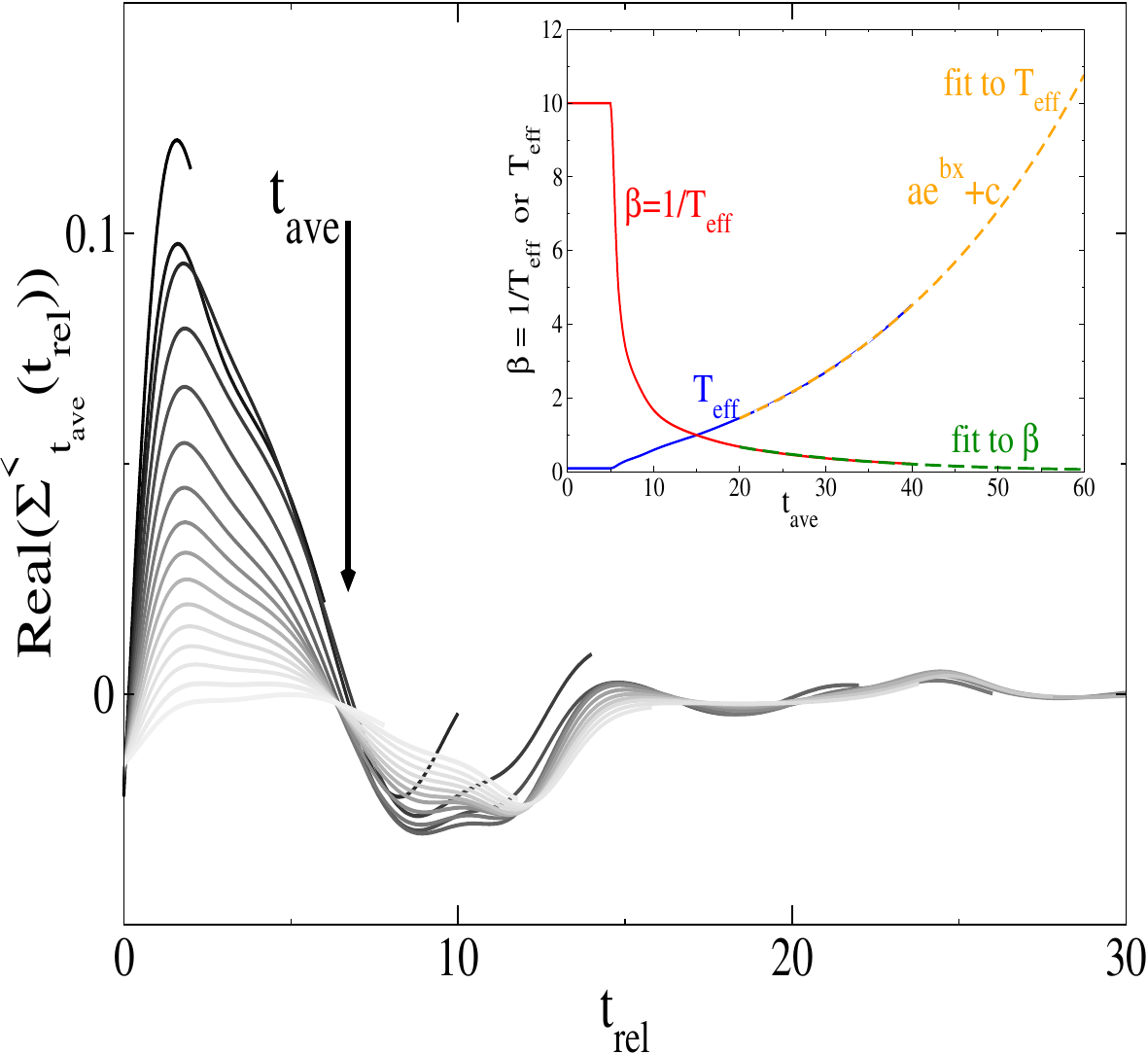}
\caption{(Color online) For parameters $E=0.5$ and $U=1.5$, the real part of the lesser self-energy as a function of relative time for successive average times. The function is odd and here, only the the positive $t_{rel}$ half is shown. A greyscale is used with darker shades indicating earlier times. Note that the deviation in the tails of the self-energy from the monotonic decay are due to one of $t$ or $t'$ being a time with the field \textit{off} while the other has the field \textit{on}. The inset shows the evolution of the effective temperature towards infinity or of its inverse towards zero. The temperature is well matched by a monotonic fit.
} 
\label{fig:SigmaLesserRealDecay_t}
\end{center}
\end{figure}

The Dyson equation for the nonequilibrium Green's function is analogous to that of the equilibrium problem. Applying the so-called Langreth rules to this produces an equation for each of the Green's functions. For $G^<$, we get:
\begin{equation}
 G^< = G_0^< + G_0^< \Sigma^A G^A +  G_0^R \Sigma^< G^A +  G_0^R \Sigma^R G^<. 
 \label{eq:DysonLesser}
\end{equation}
Since Eq.~(\ref{eq:retardedLesserParticleHole}) also holds for the noninteracting Green's function, one can infer that it holds for the self-energies as well. This is illustrated for the imaginary part of the self-energies in Fig.~(\ref{fig:SigmaRetardedSS_Eq_t}). The figure shows the calculated imaginary parts in real time of the lesser self-energy and the retarded self-energy in the steady-state calculation. The imaginary part of the lesser self-energy (black) is half of that of the retarded self-energy (red) which is well matched by the steady state value within small numerical discrepancies. The steady state data is obtained from a nonequilibrium steady state calculation performed in frequency space in a finite frequency window and then Fourier transforming to time while the transient is obtained by extrapolating to the continuum the finite time calculation for 3 commensurate time grids. The figure includes, as a reference, the imaginary part of the equilibrium retarded self-energy.

The evolution of the real part of the lesser self-energy, just like that of the Green's function, towards its vanishing value at infinite temperature is then a process, along with the time evolution of other physical quantities, that can be used to characterize the  relaxation of the system. Fig.~(\ref{fig:SigmaLesserRealDecay_t}) shows the real part of the lesser self-energy as a function of relative time for the parameters of interest ($E=0.5$ and $U=1.5$) in the monotonic thermalization scenario\cite{thermalization}. The figure uses a greyscale with darker shades indicating earlier average times. Note that the tails of the Green's functions at early average times deviate from the general monotonic decrease. This can be ascribed to the mixture, in Wigner coordinates, between the field \textit{on} and the field \textit{off} times for $t$ and $t'$. The inset shows the calculated effective temperature as a function of time and a monotonic fit that closely matches the calculated data.

Another key aspect of the monotonic thermalization scenario is that, as the system approaches its infinite temperature state, it evolves through consecutive quasi-thermal states satisfying the fluctuation-dissipation theorem\cite{NoneqFDT_Frontiers}. In terms of Green's function, the fluctuation dissipation theorem can be expressed as: 
\begin{equation}
 G^<(\omega) = -2 i f_T(\omega) \mathrm{Im}[G^R(\omega)]
 \label{eq:FDT_G_1}
\end{equation}
or equivalently:
\begin{equation}
 G^<(\omega) = - i f_T(\omega) [G^R(\omega) - G^A(\omega)].
 \label{eq:FDT_G_2}
\end{equation}
Where $f_T(\omega)$ is the Fermi-Dirac distribution at the time-dependent effective temperature $T$. To evaluate the effective temperature, we evaluate the time-dependent total energy of the system by adding to the initial equilibrium energy, the Joule heating contribution, $\int_0^t \mathbf{J}(\bar{t}) \cdot \mathbf{E} \; d\bar{t}$, where $\mathbf{J}(t)$ is the current as a function of time. This total energy is then used with the equilibrium \textit{energy versus temperature} data to infer the effective temperature of the nonequilibrium system at time $t$.

\begin{figure}[htbp]
\begin{center}
\includegraphics*[width=8.50cm, height=7.0cm]{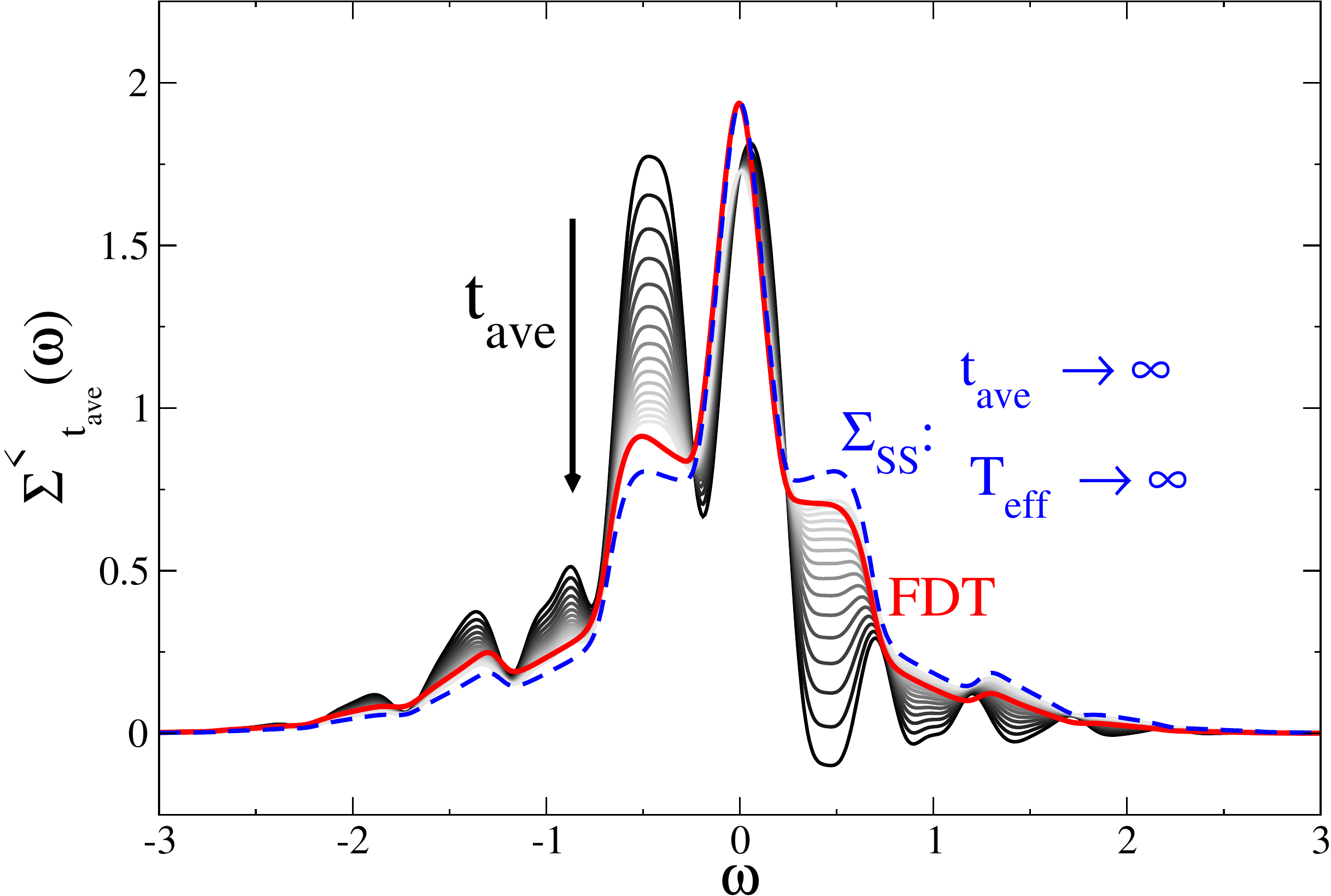}
\caption{(Color online) Fluctuation-dissipation theorem: Lesser self-energy as a function of frequency at successive average times ($E=0.5, U=1.5$). A greyscale is used with the darker shades indicating earlier times. The red curve represents the lesser self-energy obtained by applying the fluctuation-dissipation theorem with the steady-state retarded self-energy at the latest available average times. The dashed blue line shows the infinite-temperature steady-state lesser self-energy.
} 
\label{fig:SigmaLesserWCompareFDT}
\end{center}
\end{figure}

Combining the Dyson equation for the lesser Green's function in Eq.~(\ref{eq:DysonLesser})  with the fluctuation-dissipation theorem in Eq.~(\ref{eq:FDT_G_2}) for $G^<$ and $G^<_0$, one readily obtains:
\begin{equation}
 \Sigma^<(\omega) = - i f_T(\omega) [\Sigma^R(\omega) - \Sigma^A(\omega)]
 \label{eq:FDT_Sigma_1}
\end{equation}
or equivalently,
\begin{equation}
 \Sigma^<(\omega) = -2 i f_T(\omega) \mathrm{Im} [\Sigma^R(\omega)].
 \label{eq:FDT_Sigma_2}
\end{equation}

Thus, clearly, if the Green's function satisfies the fluctuation-dissipation theorem, the self-energy also does. This fluctuation-dissipation theorem for the self-energy is illustrated in Fig.~(\ref{fig:SigmaLesserWCompareFDT}) where the evolution of the imaginary part of the lesser self-energy (in frequency space) towards its steady-state value at infinite temperature is presented. The greyscale shows the transient calculation with darker shades indicating earlier times. The red line shows the fluctuation dissipation result at the latest calculated transient time, while the dashed blue curve shows the infinite-temperature steady-state result. Note the agreement between the fluctuation-dissipation theorem result and the transient calculation at the corresponding average time.

\begin{figure}[htbp]
\includegraphics*[width=8.50cm, height=7.0cm]{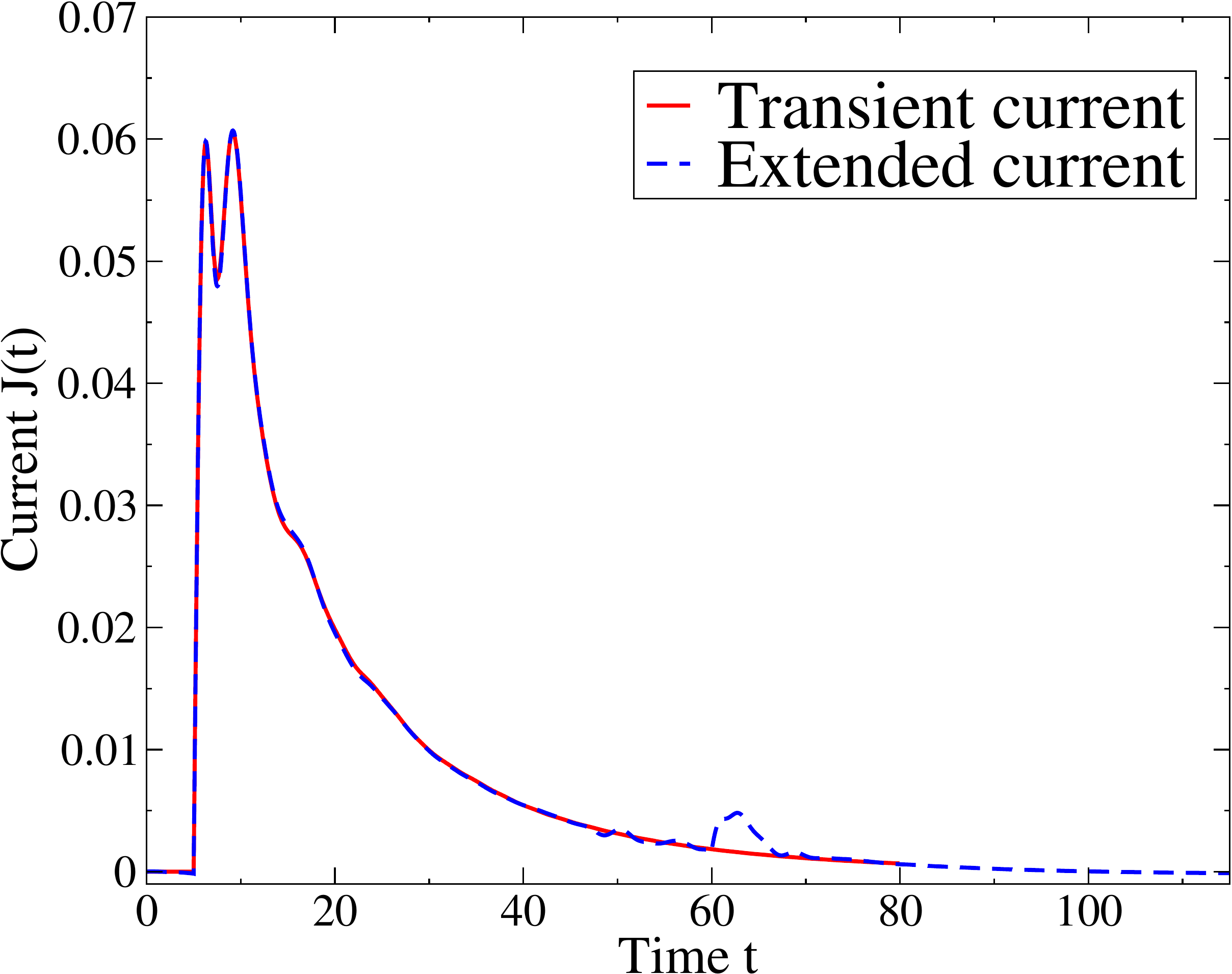}
\caption{(Color online) Current as a function of time for $E=0.5, U=1.5$. The solid red line represents the transient calculation for a nonequilibrium DMFT extending up to $t_{max} = 80$. The dashed blue line shows the current obtained by extending the transient self-energy out to $t_{max}^{New}=115$ where the current is near its steady-state zero value. The extended current exhibits a ripple around the patching time between the short transient result and the fluctuation dissipation theorem extended result. This ripple is due to the small discontinuity between the two data sets. However this extended current clearly overlaps with the tail of the transient DMFT calculation. More robust blending between these solutions should reduce the spurious oscillation near the patch time.
} 
\label{fig:extendedCurrent}
\end{figure}

The fluctuation-dissipation theorem satisfied by the system as discussed above, coupled with the monotonic evolution of the temperature well matched by the fit to the transient data can allow us to extend the reach of the calculation beyond the initial transient. To this end, the fit to the transient effective temperature curve allows us to obtain the temperature as a function of time starting from a time $t_{patch}$, which is after the initial nontrivial transient and after the system has settled into its monotonic evolution. With the fit, the temperature versus time data can be extended beyond $t_{max}$ and up to an arbitrary new average time $t_{max}^{New}$. Next, using the fluctuation-dissipation theorem for the self-energy in Eq.~(\ref{eq:FDT_Sigma_2}) as a function of frequency, we can obtain frequency-dependent lesser self-energies for average times up to $t_{max}^{New}$. If we then perform a Fourier transform on this result, we get $\Sigma^<(t_{ave}, t_{rel})$ covering the larger square blue box of Fig.~(\ref{fig:Keldysh2Wigner}). The data can be converted as needed back to the $(t,t')$ contour time coordinates. The imaginary-time vertical spur is unchanged while the mixed time lesser self-energy has essentially decayed to zero before the time $t_{patch}$. Coupled with the knowledge of the retarded self-energy that is only constrained by causality, the knowledge of the lesser self-energy as a function of time allows us, following this procedure, to construct the full contour-ordered self-energy. Since the DMFT formalism assumes a local self-energy, the lattice Green's function can now be calculated using in the Dyson equation, this extended self-energy up to much later times than the original transient calculation. This accordingly enables the extraction of other relevant lattice quantities.

While there are small deviations at the patching line between the transient DMFT calculation and the fluctuation dissipation theorem extended data, we note the satisfactory agreement between the data obtained by the two methods. The self-energy is thus extended in average time well beyond the original calculated transient. Fig.~(\ref{fig:extendedCurrent}) shows the electric current as a function of time for this system ($E=0.5$ and $U=1.5$). The extended current is obtained by following the procedure described above with $t_{patch} = 30$ on the transient DMFT and extending the self-energies out to $t_{max}^{New}=115$. At infinite temperature, the current is expected to vanish as all momentum points are equally likely, i.e. electrons are equally likely to move in all directions. A transient calculation, performed at a higher computational cost, gives the current up to a maximal time $t_{max} = 80$. Our extrapolation scheme is shown to enable calculation of the current up to times where it is close to its steady value. Note that the ripple in the extended current around time $t_{patch}$ is a numerical artifact that can be ascribed to the discontinuity at the patching line between the two numerical methods, the transient DMFT self-energies and the extrapolated FDT self-energies. Additionally, although the mixed-time self-energies have decayed to small values at the patching time, they numerically have finite values ($\sim 10^{-4}$). Nevertheless, the transient calculation with a larger maximum time $t_{max} = 80$ is shown to indeed overlap in its tail with the extrapolated current that is extended to much later times. It should be clear that this approach can be applied to other systems where the long-time-limiting behavior is known and, for which, the approach to the steady state has similar predictive behavior as we saw with this example.

\section{Conclusion}
\label{sec: Conclusion}
We have used an analysis of the impurity self-energy in the transient and in the steady-state nonequilibrium DMFT solution for a system described in equilibrium by the Falicov-Kimball model. The model describes a Fermi-Fermi heavy-light mixture and exhibits a variety of nontrivial behaviors including that of the Mott insulator transition. Our analysis has in particular focused on the case of a monotonic evolution towards an infinite temperature thermal state when the system, initially in equilibrium at a temperature $T=1/\beta=0.1$, is suddenly placed under the influence of a DC electric field along the diagonal of an infinite-dimensional hypercubic lattice. Here, just like the Green's function, the lesser self-energy, after an initial nontrivial stage in its transient, evolves towards its steady state by going through successive values obeying the fluctuation-dissipation theorem at the associated effective temperature. By fitting a monotonic function to the curve of the effective temperature as a function of average time in the later stages of the transient, we are able to employ the fluctuation-dissipation theorem to extend the characterization of the system well beyond the initial transient calculation at a minimal computational cost. Such a scheme to bridge the gap between a transient calculation and a calculation directly in the steady state is of significant value for studying the full dynamics of this nonequilibrium quantum system. It will be worth exploring for other scenarios of the relaxation of a field-driven systems and for other nonequilibrium systems. \\

\section*{Acknowledgments} 

HFF is supported by the National Science Foundation under Grant No. PHY-2014023.
JKF was supported by the Department of Energy, Office of Basic Energy Sciences, Division of Materials Sciences and Engineering under Contract No. DE-FG02-08ER46542. JKF was also supported by the McDevitt bequest at Georgetown. AFK acknowledges support by the National Science Foundation under Grant No. DMR-1752713.

\end{document}